\documentclass[conference,a4paper]{IEEEtran}
\IEEEoverridecommandlockouts
\usepackage{cite}
\usepackage{indentfirst}
\usepackage{multirow}
\usepackage{booktabs}
\usepackage{makecell}
\usepackage{float}
\usepackage{amsmath,amssymb,amsfonts}
\usepackage{algorithmic}
\usepackage{graphicx}
\usepackage{graphicx,subfigure}
\graphicspath{{./figures/}}
\usepackage{textcomp}
\usepackage{xcolor}
\usepackage{subfigure}
\usepackage{booktabs}
\usepackage{bm}
\usepackage{amsthm}
\usepackage{epstopdf}
\def\BibTeX{{\rm B\kern-.05em{\sc i\kern-.025em b}\kern-.08em
		T\kern-.1667em\lower.7ex\hbox{E}\kern-.125emX}}

\begin{document}
	\bibstyle{IEEEtran}
	\title{Towards Native Intelligence: 6G-LLM Trained with Reinforcement Learning from NDT Feedback}
	
	\author{
		\IEEEauthorblockN{Zhuoran~Xiao,
			Tao~Tao,
			Chenhui~Ye,
			Yunbo~Hu,
			Yijia~Feng,
			Tianyu~Jiao,
			and Liyu~Cai
		}
		
		\IEEEauthorblockA{Nokia Bell Labs, Shanghai, China \\
			E-mails: \{zhuoran.xiao, tao.b.tao, chenhui.a.ye, yunbo.hu, yijia.feng, tianyu.jiao, liyu.cai\}@nokia-sbell.com}
	}
	\maketitle
	
	\begin{abstract}
		Owing to its comprehensive understanding of upper-layer application requirements and the capabilities of practical communication systems, the 6G-LLM (6G domain large language model) offers a promising pathway toward realizing network native intelligence. Serving as the system orchestrator, the 6G-LLM drives a paradigm shift that fundamentally departs from existing rule-based approaches, which primarily rely on modular, experience-driven optimization. By contrast, the 6G-LLM substantially enhances network flexibility and adaptability. Nevertheless, current efforts to construct 6G-LLMs are constrained by their reliance on large-scale, meticulously curated, human-authored corpora, which are impractical to obtain in real-world scenarios. Moreover, purely offline-trained models lack the capacity for continual self-improvement, limiting their ability to adapt to the highly dynamic requirements of wireless communication environments. To overcome these limitations, we propose a novel training paradigm termed RLDTF (Reinforcement Learning from Digital Twin Feedback) for 6G-LLMs. This framework leverages network digital twins to generate reward signals based on orchestration outcomes, while employing reinforcement learning to guide the model toward optimal decision-making dynamically. Furthermore, we introduce a weighted token mechanism to improve output accuracy. Comprehensive experimental results demonstrate that our proposed framework significantly outperforms state-of-the-art baselines in orchestration accuracy and solution optimality.
	\end{abstract}
	
	
	\section{Introduction} \label{intro}
	With the introduction of emerging features and functionalities such as artificial intelligence (AI) and sensing, and the support for novel application scenarios like edge computing service, 6G systems are expected to exhibit a substantially higher degree of complexity. Traditional systems, which rely on fixed and relatively simple workflows with module-by-module optimization and orchestration, will no longer be sufficient to fully unlock the network's potential. Furthermore, with the integration of AI-empowered communication modules, the functions of future networks can be flexibly realized through multiple alternative pathways. As a result, understanding the capabilities and characteristics of each tool within the system toolbox and achieving their optimal orchestration will become a central focus of network evolution. In addition, current networks cannot comprehend upstream service content, failing to deliver fine-grained configurations tailored to specific service requirements. These challenges collectively drive wireless systems toward an evolution characterized by native intelligence.
	
	Due to their strong capabilities in cognition, comprehension, planning, and reflection across general-purpose tasks, large models are regarded as the most promising and feasible approach to addressing the aforementioned challenges. However, the construction objectives of large models in the 6G domain fundamentally differ from those of general-purpose models, which raises new challenges. 
	
	First, domain-specific large models must incorporate additional communication knowledge while preserving their core general capabilities. Second, while the ultimate goal of general-purpose large models is to ensure that the output token distribution is reasonable, they often struggle to produce exact outputs. In contrast, the objective of 6G-LLM is not only to generate reasonable responses but also to provide precise numerical parameters. Lastly, while general-purpose large models benefit from abundant accessible corpora, training 6G-LLM relies heavily on large-scale, high-quality expert corpora. However, such corpora are extremely difficult to obtain from real-world systems, and their quality cannot be directly assessed through manual inspection.
	
	\begin{figure*}[htb!]
		\centering
		\includegraphics[width=0.95\textwidth]{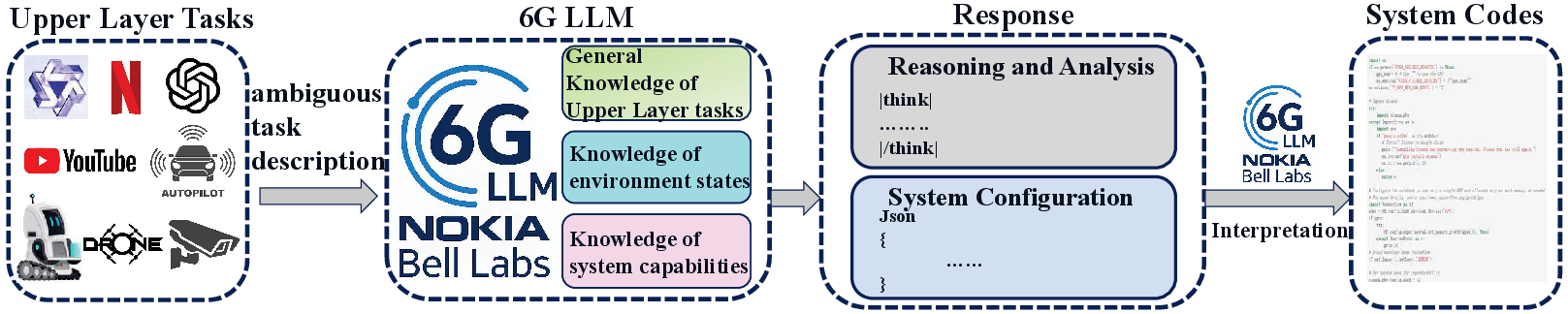}
		\vspace{-.2cm}
		\caption{System diagram of the proposed native intelligence communication system enabled by 6G-LLMs.}
		\label{system_diagram}
		\vspace{-.5cm}
	\end{figure*}
	Several studies have proposed the vision of realizing network intelligence through large models. The potential benefits and open issues of employing large language models as managers for 6G networks are discussed in \cite{10648594,10700707,10638533}. Authors in \cite{11097898} attempted to build a telecom-specific LLM for understanding domain-specific documents. Leveraging multi-modal LLM for solving specific physical layer tasks is discussed in \cite{11029511}. However, most existing studies remain at the visionary stage or are merely based on general foundation models, without proposing concrete training methodologies.
	
	In our prior work \cite{11101226}, we addressed this challenge to some extent by injecting domain knowledge and enabling the model to recognize and understand the system’s functionalities to orchestrate them effectively. Nevertheless, this effort remains far from perfect. First, teaching the system to orchestrate correctly requires a large volume of training samples crafted by experienced engineers, which is impractical in real-world scenarios. Moreover, human experts can only provide reasonable solutions rather than optimal ones. Besides, due to the lack of self-evolution capabilities, networks trained in this manner cannot adapt to the highly dynamic and complex nature of practical communication environments.
	
	This paper proposes a novel training paradigm for 6G-LLMs, termed Reinforcement Learning from Digital Twin Feedback (RLDTF). In this framework, a pre-trained and fine-tuned 6G-LLM receives feedback on network orchestration quality from a digital twin that simulates the physical transmission process. It iteratively refines its outputs based on this feedback to achieve transmission optimization. Guided by this principle, we introduce a new mechanism that leverages system-level digital twins to provide feedback signals for large model training. Furthermore, we design a novel reinforcement learning algorithm tailored to the unique optimization objectives inherent to communication systems. Finally, recognizing the fundamental differences between 6G-LLMs and general-purpose LLMs, particularly the decisive influence of key system configuration parameters, we propose a weighted token advantage mechanism for reinforcement learning to enhance the overall performance.
	
	The remainder of this paper is structured as follows. Section \ref{Problem} presents the system model and the task objective. The training process, algorithm, and motivation behind it are discussed in Section \ref{process_design}. Section \ref{experiments} details the experimental setup and reports the results, providing a comparative evaluation of the proposed approach against existing state-of-the-art (SOTA) models from multiple perspectives. Finally, Section \ref{conclusion} summarizes the main conclusions of this work.
	
	\section{System Model And Task Description} \label{Problem}
	As illustrated in Fig. \ref{system_diagram}, this paper investigates a novel communication system powered by 6G-LLMs, in which 6G-LLMs serve three primary roles. First, 6G-LLMs act as the bridge between upper-layer task intents and the transmission workflow orchestration. Existing communication systems control air interface configurations only through coarse-grained service classification, making it difficult to adjust the system finely according to particular service requirements. In contrast, 6G-LLMs function as a natural knowledge base capable of understanding each specific application's QoS (quality of service) requirements for the transmission system, thereby enabling precise translation from service intent to system-level QoS configurations.
	
	Second, the trained 6G-LLMs possess comprehensive knowledge about the communication equipment, including hardware parameters, real-time operational status, and detailed descriptions of each API in their toolkit. Therefore, by jointly considering the transmission QoS requirements associated with high-level services and the capabilities supported by the system devices, the 6G-LLMs can perform workflow orchestration and parameter configuration to optimize specific transmission tasks. Moreover, by enhancing the reasoning capabilities of the 6G-LLMs through training, the model acquires self-reflective ability, enabling it to evaluate and refine its own outputs, thereby ensuring reliability and robustness.
	
	Finally, the 6G-LLMs can translate workflow orchestration and API arguments described in structured natural language into executable code directly deployed in the practical system. This capability enables dynamic and flexible reconfiguration of system configuration on demand, thereby realizing a high level of autonomous intelligence in communication systems.
	
	\begin{figure*}[htb!]
		\centering
		\includegraphics[width=1\textwidth]{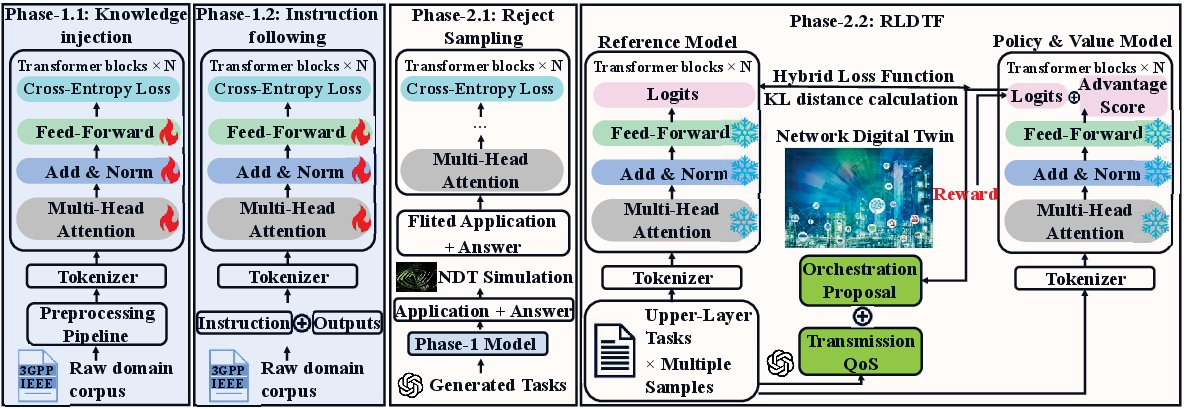}
		\vspace{-.8cm}
		\caption{The training process the proposed 6G-LLMs.}
		\label{training_process}
		\vspace{-.7cm}
	\end{figure*}
	
	\section{Training Process Design and Motivation} \label{process_design}
	As shown in Fig. \ref{training_process}, the training process of the 6G-LLM consists of two stages and four steps, encompassing multiple phases including pre-training, fine-tuning, and reinforcement learning. This section will present the relevant procedures and algorithms in chronological order according to the training pipeline.
	\subsection{Pre-Training and Fine-Tuning}
	The primary objective of 6G-LLM pre-training is to inject domain-specific knowledge. We collect as many technical documents as possible from the 3GPP FTP, IEEE publications, and company-internal proprietary repositories. Subsequently, the raw documents are processed and refined into a high-purity domain-specific corpus through manually designed corpus cleaning scripts. Additionally, to mitigate catastrophic forgetting, we mix a proportion of open-source general-purpose corpora with the domain-specific data. Details regarding corpus preparation and model training can be referred to our prior work \cite{11101226}. It is worth mentioning that, different from prior work, we employed full-parameter training to enhance the model’s learning capacity and knowledge integration due to the significantly expanded scale of the collected domain corpus.
	
	Typically, a model's general instruction-following capability may degrade after pre-training. To address this, we introduce a domain-specific instruction-following training phase. Using a third-party model and prompt engineering, we converted a subset of domain-specific corpora into question-answer (Q\&A) format and mixed them with a proportion of general-purpose data. The model is then fine-tuned with full-parameter updates, employing a gradually decaying learning rate. This results in a first-stage model that retains strong general capabilities while incorporating rich domain knowledge and enhanced question-answering performance. It is worth emphasizing that we incorporated a substantial amount of reasoning data into the dataset. This shift encourages the model to shift from simple direct responses to a \textit{think-reflect-answer} generation mode. As a result, the context length of the model's responses increases, enabling more extensive retrieval and utilization of domain-specific knowledge, further improving answer accuracy.
	\subsection{Reject Sampling}
	For reinforcement training of large language models, the initial model performance is crucial for training convergence and robustness. Excessive negative feedback at the initial training stage can severely hinder convergence. To address this issue, we propose a rejection sampling-based approach that efficiently optimizes the initial model state by selectively training on high-quality positive samples. 
	
	We first generate a large number of upper-layer task descriptions using a third-party LLM, each paired with the minimum transmission requirements, such as latency, throughput, and bit error rate, necessary to fulfill the corresponding task. It is worth mentioning that, in a practical system, such data should ideally be collected from historical transmission records. Subsequently, we manually filter out unreasonable task descriptions to construct a high-quality seed corpus. Using this seed corpus, the first stage 6G-LLM generates system workflow orchestrations and parameter configurations for the APIs based on the given service QoS demands. The model is then invoked once more to translate these orchestrations into executable code for the digital twin simulation system, where the actual QoS metrics of the simulated transmission are evaluated. Finally, only those input-output pairs that successfully meet the target QoS requirements are selected to form the training dataset for subsequent model refinement.
	
	Notably, this process can be iterated multiple times, progressively improving model performance until the success rate of valid orchestrations stops growing. This approach effectively replaces the typically unstable initial exploration phase in reinforcement learning with a guided, positive-sample-based training strategy, stabilizing early-stage learning and accelerating convergence.
	\subsection{Reinforcement Learning from NDT Feedback}
	\subsubsection{Problem Formulation}
	We formulate the training of the 6G-LLM as a sequential decision-making process under a reinforcement learning framework. At each generation step $t$, the state $s_t$ consists of two components. One is the upper-layer task description $\mathbf{u}$. The other is the sequence of tokens already generated $\mathbf{a}_{<t}=(a_1,a_2,...,a_{t-1})$. The action space is the tokenizer's vocabulary $\mathbf{\nu}$, and the action $a_t\in \mathbf{\nu}$ corresponds to the selection of the next token in the output sequence. The policy $\pi_\theta (a_t|s_t)$ is implemented by the 6G-LLM parameterized by $\theta$, which outputs a probability distribution over the vocabulary.
	
	The generation process continues until an end-of-sequence token is produced, yielding a complete orchestration command $\mathbf{a}=(a_1,...,a_T)$. This sequence is then interpreted to the executable code of the digital twin platform, denoted by $\mathbf{c}=f_{\theta}(\mathbf{a})$. The simulated results that contain the transmission QoS conducted by NDT are denoted by $\mathbf{q}=\mathbf{NDT}(\mathbf{c})$. The reward is calculated by a pre-defined function $\mathbf{R}(\cdot)$, which can be written as $r=\mathbf{R}(\mathbf{u},\mathbf{a})$. Since the reward is only available after the full sequence is generated, the learning objective is to find the optimal policy parameters that maximize the expected cumulative reward:
	\begin{equation}
		\theta^* = \arg\max_{\theta} \, \mathbb{E}_{\mathbf{a} \sim \pi_\theta(\cdot \mid \mathbf{u})} \big[ \mathbf{R}(\mathbf{u}, \mathbf{a}) \big]
	\end{equation}
	\subsubsection{Reward Function Design}
	The transmission QoS conducted by NDT includes three metrics, including processing delay $\hat{delay}$, throughput $\hat{thr}$, and bit error rate (BER) $\hat{BER}$, which can be denoted as $\mathbf{q}=(\hat{del},\hat{thr},\hat{BER})$. Correspondingly, the original upper-layer task description specifies the required targets for these three QoS metrics, which are maximum allowable processing delay $del$, minimum required throughput $thr$, and maximum tolerable bit error rate $BER$. 
	
	We design the reward function to encourage the 6G-LLM to generate orchestrations that satisfy fundamental QoS requirements while minimizing communication resource wastage. Therefore, when the generated configuration satisfies the basic QoS requirements, a penalty proportional to the degree of communication resource wastage is subtracted from a base positive reward. Conversely, to ensure a smooth reward landscape, when the QoS requirements are violated, a proximity-based bonus, reflecting how close the achieved performance is to the target, is added to a base negative penalty. The reward function can be written as
	\begin{equation}
		\mathbf{R} = 
		\begin{cases}
			R_{\text{base}}^{+}-w_{1}^{+}(\hat{thr}-thr)-w_{2}^{+}(del\\-\hat{del})-w_{3}^{+}(BER-\hat{BER} ), & \text{if QoS satisfied}, \\
			R_{\text{base}}^{-} + w_{1}^{-}sig(\frac{1}{|\hat{thr}-thr|})+\\w_{2}^{-}sig(\frac{1}{|\hat{del}-del|})+\\w_{3}^{-}sig(\frac{1}{|BER-\hat{BER}|}), & \text{otherwise},
		\end{cases}
	\end{equation}
	where $R_{\text{base}}^{+}=1$ and $R_{\text{base}}^{-}=-1$. $w_{1}^{+}$, $w_{2}^{+}$, $w_{3}^{+}$, $w_{1}^{-}$, $w_{2}^{-}$, and $w_{3}^{-}$ are weight coefficient. $sig(\cdot)$ denotes the Sigmoid function. $|\cdot|$ is the absolute value function.
	\subsubsection{Token Importance Estimation via Perturbation-Based Reward Sensitivity}
	We propose a perturbation-based sensitivity analysis to identify tokens that most significantly influence the reward and decide the token advantage weight for reinforcement training. The core idea is to perturb some tokens and estimate the change to the reward function. Those tokens that lead to a large change are deemed critical.
	
	Given a prompt $\mathbf{u}$, the policy \( \pi_\theta \) generates a baseline orchestration sequence $\mathbf{a} = (a_1, a_2, \dots, a_T) \sim \pi_\theta(\cdot \mid \mathbf{u})$, which yields a scalar reward $R_0$. For each token position $t \in \{1, \dots, T\}$, we generate $N$ perturbed sequences $ \{\tilde{\mathbf{a}}^{(n,t)}\}_{n=1}^N$, where only the $t$-th token is modified as
	\begin{equation}
		\tilde{a}^{(n,t)}_{t'} =
		\begin{cases}
			\text{Perturb}(a_t), & \text{if } t' = t, \\
			a_{t'}, & \text{otherwise},
		\end{cases}
	\end{equation}
	where $\text{Perturb}(\cdot)$ is a perturbation function that randomly replaces this token with one that appears in the response or directly removes it. Each perturbed sequence is executed to obtain a reward $R^{(n,t)} = \mathbf{R}(\mathbf{u}, \tilde{\mathbf{a}}^{(n,t)})$.
	The reward sensitivity of token $t$ is defined as the average absolute reward deviation caused by its perturbation
	\begin{equation}
		S_t = \frac{1}{N} \sum_{N=1}^N \left| R_0 - R^{(n,t)} \right|.
	\end{equation}
	The final token weight is calculated by 
	\begin{equation}
		w_t = 1 + \alpha \cdot \frac{S_t}{\max_{t'} S_{t'} + \lambda},
	\end{equation}
	where $\lambda > 0$ is a small constant for numerical stability. $\alpha \geq 0$ is a scaling hyperparameter that controls the weighting strength. It is worth mentioning that token weighting is applied only to tokens whose reward sensitivity exceeds a predefined threshold in practice.
	\subsubsection{Reinforcement Training Algorithm}
	During the reinforcement learning phase, we deploy three models. First is the action model, which shares the same architecture as the foundation model and is initialized using the model obtained after reject sampling. Second is the critic model. Since the rewards provide scores for the entire generated token sequence, the critic model must further estimate each token's advantage to compute the loss function. The critic model $V_\phi(\cdot)$ retains the backbone of the foundation model but replaces the original language modeling (LM) head with a value head that outputs a scalar value. The critic is trained jointly with the action model. Third is the reference model, which is initialized in the same manner as the action model. However, throughout training, its parameters are updated only by copying those of the action model at checkpoint-saving intervals, and remain frozen otherwise. This reference model serves as a stable baseline for policy updates, preventing excessive deviation in the learning direction and ensuring training robustness.

	The total loss function consists of three components. One is the policy loss with token weighting, which can be written as
	\begin{align}
		&\mathcal{L}^{\text{policy}}(\theta) =\notag \\& -\mathbb{E}_{a \sim \pi_{\theta_{\text{old}}}} \left[ \sum_{t=1}^{T} w_t \cdot \min\left( r_t(\theta) A_t,\ \text{clip}\big(r_t(\theta), \epsilon \big) A_t \right) \right],
	\end{align}
	where \( r_t(\theta) = \frac{\pi_\theta(a_t \mid s_t)}{\pi_{\theta_{\text{old}}}(a_t \mid s_t)} \). $A_t = \mathbf{R}(\mathbf{u}, \mathbf{a}) - V_\phi(s_t)$ is the approximated token advantage. $\epsilon$ is a hyper-parameter. \( w_t \) is the token weight. The second is the value function loss, which calculates the squared error between the observed return and the value estimated, which can be written as 
	\begin{equation}
		\mathcal{L}^{\text{value}}(\phi) = \mathbb{E} \left[ \sum_{t=1}^{T} \left( \mathbf{R} - V_\phi(s_t) \right)^2 \right].
	\end{equation}
	The third is the entropy bonus and the KL Regularization function to encourage exploration and prevent divergence from the SFT model, which is
	\begin{align}
		&\mathcal{L}^{\text{reg}}(\theta) = \notag \\& -\beta_{\text{ent}} \cdot \mathbb{E}[\mathcal{H}(\pi_\theta(\cdot \mid s_t))] + \beta_{\text{KL}} D_{\text{KL}}\big( \pi_\theta(\cdot \mid s_t) \,\|\, \pi_{\theta_{old}}(\cdot \mid s_t) \big),
	\end{align}
	where $-\beta_{\text{ent}}$ is the entropy bonus coefficient to control the exploration. $\mathcal{H}(\cdot)$ denotes the Shannon entropy. $\beta_{\text{KL}}$ is the KL regularization coefficient. $D_{\text{KL}}(\cdot)$ denotes the KL divergence. 
	Thus, the final loss function is
	\begin{equation}
		\mathcal{L}_{\text{total}}(\theta, \phi) = 
		\mathcal{L}^{\text{policy}}(\theta) 
		+ k_1 \mathcal{L}^{\text{value}}(\phi) 
		+ k_2 \mathcal{L}^{\text{reg}}(\theta),
	\end{equation}
	where $k_1$ and $k_2$ are hyper-parameters.
	
	\section{Experiments} \label{experiments}
	\subsection{Experiments Settings}
\begin{figure}
		\centering
		\includegraphics[width=0.4\textwidth]{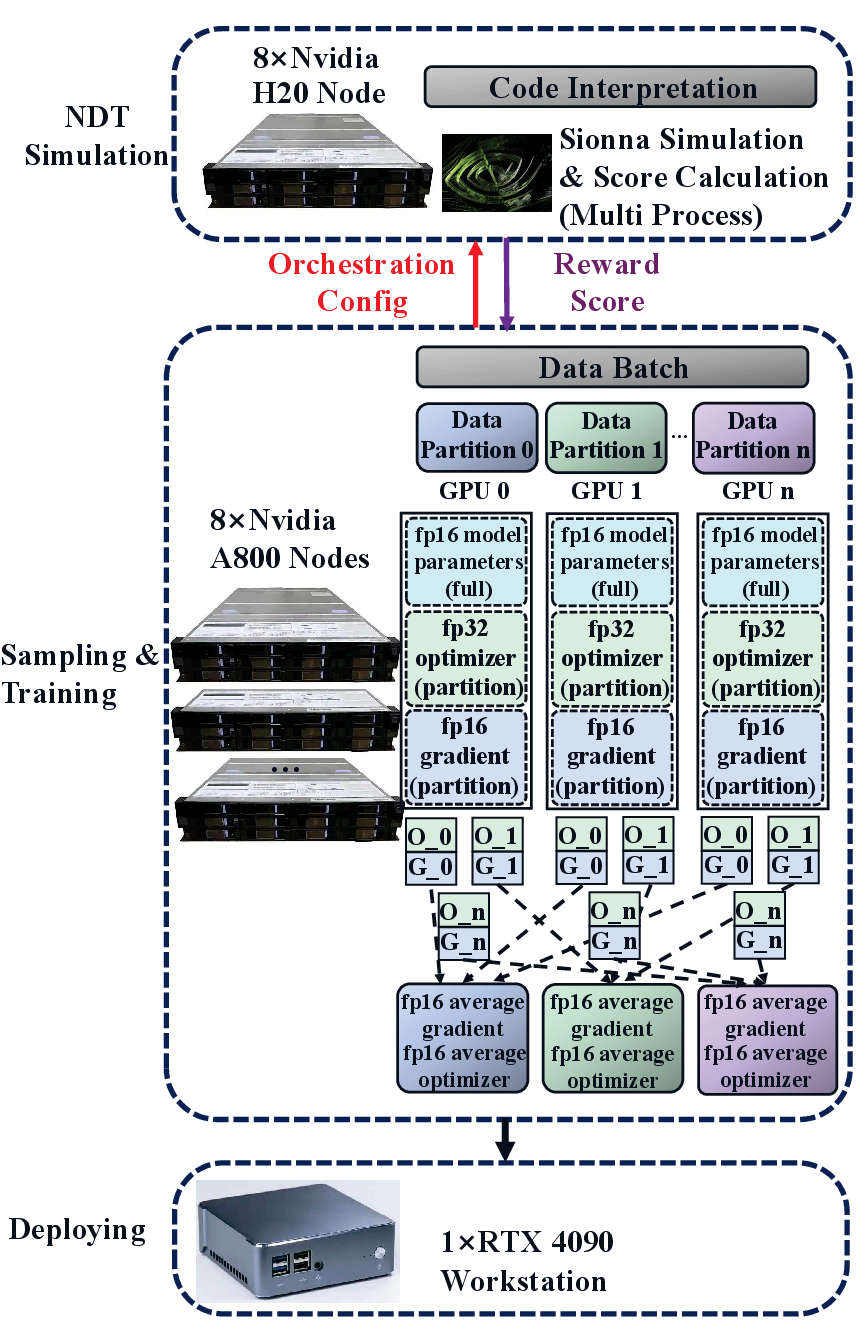}
		\vspace{-.3cm}
		\caption{Experimental framework for model training and inference.}
		\vspace{-.5cm}
		\label{experiment_framework}
	\end{figure}

	Considering practical system constraints, where large models for RAN (Radio Access Network) are typically deployed at base stations with limited computational resources, we select Qwen2.5-7B as the foundation model by balancing model capability and computational cost. As shown in Fig. \ref{experiment_framework}, we conduct model pre-training, fine-tuning, and reinforcement learning using two connected GPU servers, each equipped with 8 × NVIDIA Tesla A800 GPUs. During the reinforcement training phase, an additional GPU server with 8 × NVIDIA Tesla H20 GPUs is employed to host the LLM's inference service and the NDT simulation platform. For the digital twin platform, we adopt Nvidia Sionna to perform link-level simulations. The final trained model is lightweight enough to be deployed on edge devices with a single consumer-grade GPU.
	
	During the training process, we load model parameters in bfloat16 precision and employ the DeepSpeed ZeRO-3 framework to minimize training memory overhead. For model inference, we leverage the vLLM framework to enable dynamic parallel inference. Furthermore, to reduce training time overhead during reinforcement learning, we adopt process-level parallelism to run batch-sized digital twin simulations concurrently.
	\subsection{Dataset Generation}
	In the pre-training phase, after collecting and cleaning domain-specific corpora, we obtain approximately $30$B tokens of domain data. Furthermore, 10B tokens of general-purpose corpora are mixed. During the instruction following training step, we employ DeepSeek-R1-671B to generate approximately $150$K reasoning-intensive Q\&A pairs. Subsequently, using prompt engineering with the same model, we generated around $280$K questions covering upstream business scenarios and corresponding communication system QoS requirements. Through two rounds of rejection sampling, we curated the dataset, retaining approximately $170$K high-quality and valid instances for training.
	\subsection{Benchmarks}
	To validate the effectiveness of each training stage and its impact on overall performance, we select four comparison models as benchmarks.
	\begin{itemize}
		\item \textbf{Benchmark 1}: The raw Qwen2.5-7B-Instruct model as the foundation model.
		\item \textbf{Benchmark 2}: The raw Llama2-7B-Instruct model as the third-party model for cross-validation.
		\item \textbf{Benchmark 3}: The Phase-1 model of 6G-LLMs, to validate the necessity of domain knowledge injection.
		\item \textbf{Benchmark 4}: The refined model with reject sampling without reinforcement training.
	\end{itemize}
	
	In the performance comparison, we consider two key evaluation metrics. First is the task completion rate, which measures the proportion of queries for which the system's orchestration results satisfy the final task requirements. Second, the average score of completed tasks, which evaluates the optimality of the orchestration for successfully completed tasks. The scoring function is identical to the reward function used during reinforcement learning.
	
	\subsection{Experimental Results}
\begin{figure}
		\centering
		\includegraphics[width=0.3\textwidth]{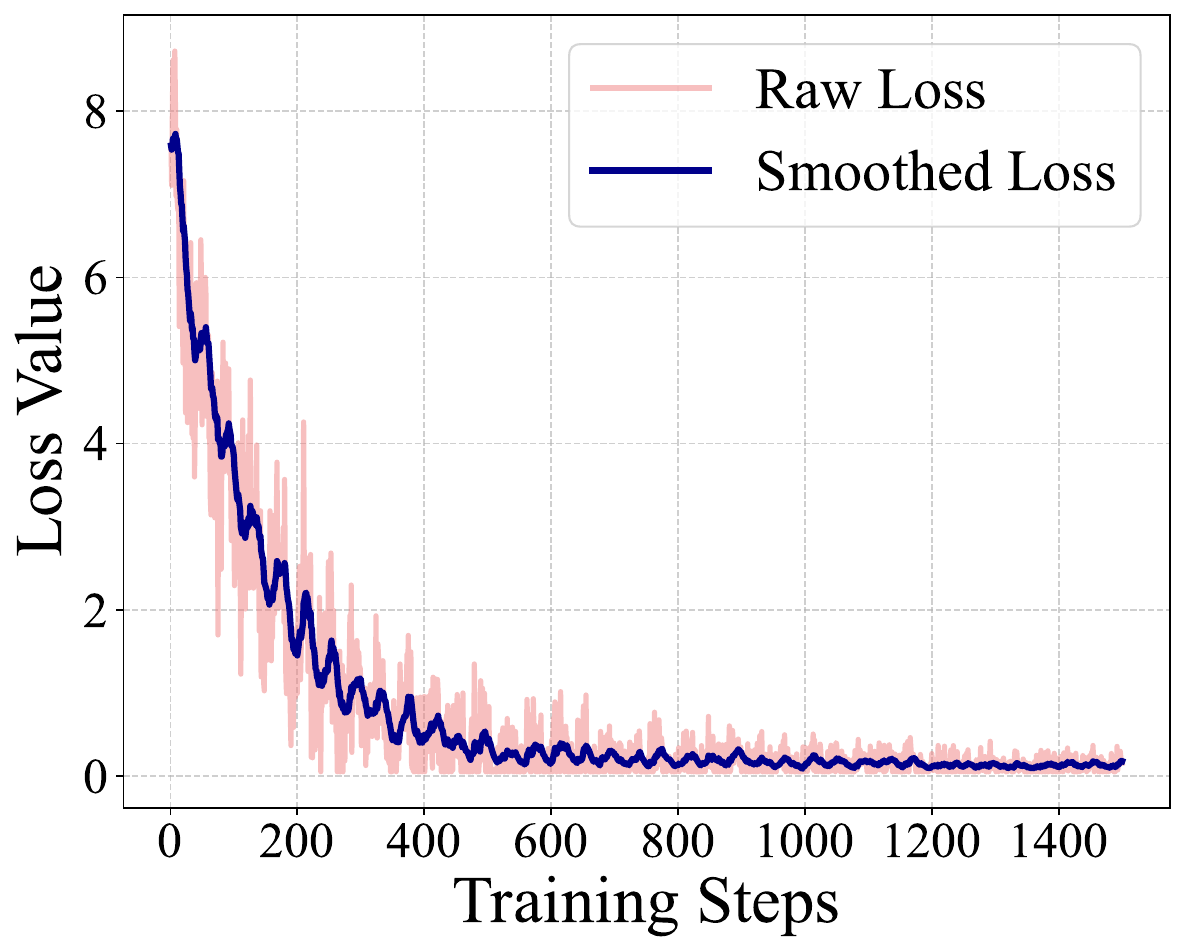}
		\vspace{-.4cm}
		\caption{Convergence curve of the loss function during reinforcement training.}
		\label{loss_curve}
		\vspace{-.5cm}
	\end{figure}

\begin{figure}
		\centering
		\includegraphics[width=0.3\textwidth]{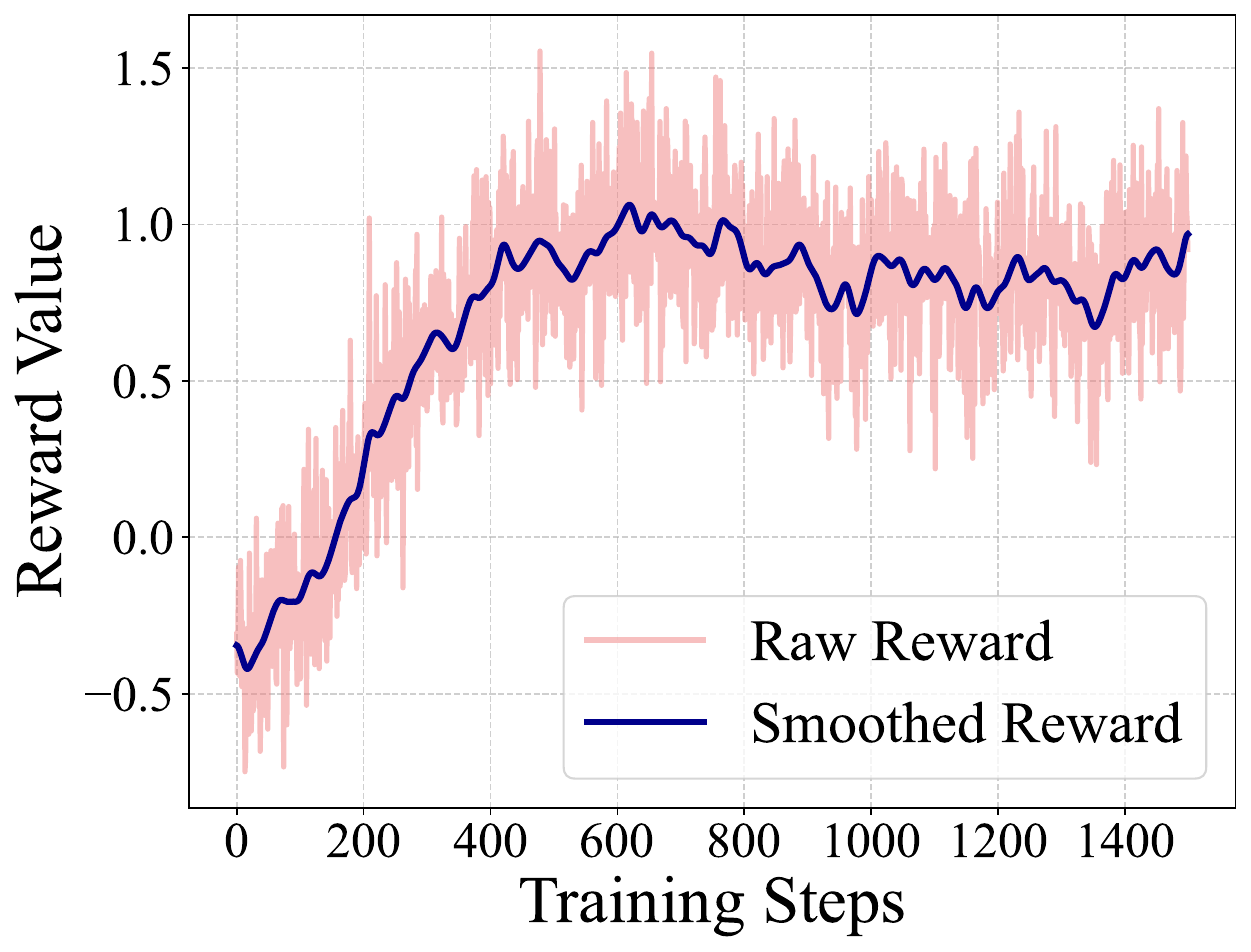}
		\vspace{-.4cm}
		\caption{Curve of the reward value during reinforcement training.}
		\label{reward_curve}
		\vspace{-.7cm}
	\end{figure}

	The loss and reward trajectories during reinforcement training are shown in Fig. \ref{loss_curve} and Fig. \ref{reward_curve}. The policy loss decreases from an initial value of about $7.5$ to below $0.5$ within the first $450$ steps, indicating rapid alignment between the LLM's response and the expected policy updates. Concurrently, the average episode reward rises steadily from approximately $-0.3$ to over $0.75$ in the same period, demonstrating effective learning. In the subsequent phase (steps $450$–$1500$), the reward stabilizes around $0.8$, suggesting that the policy has converged to a robust and consistent solution.
	
	\begin{figure}[htb!]
		\centering
		\includegraphics[width=0.35\textwidth]{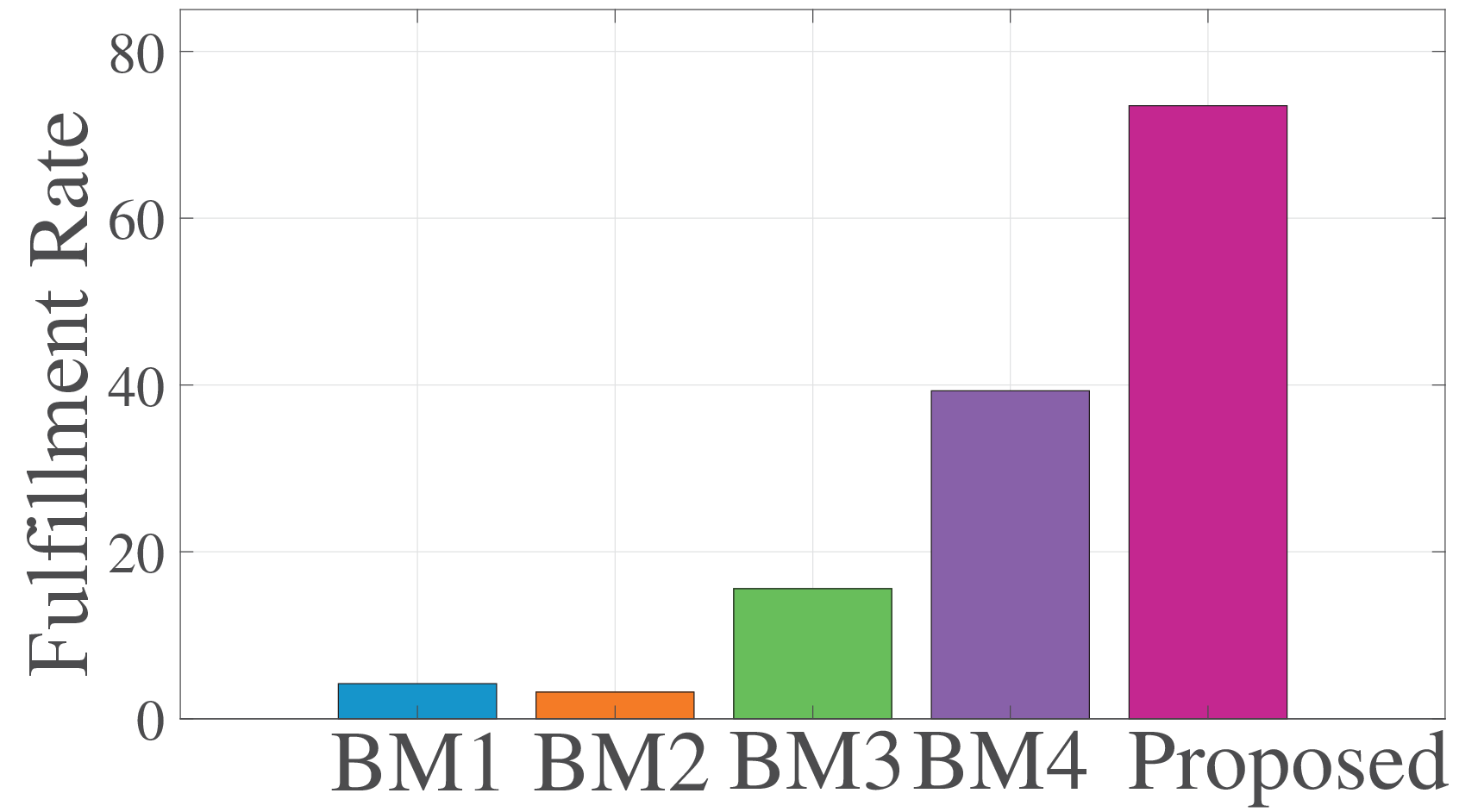}
		\vspace{-.3cm}
		\caption{Task requirement fulfillment ratio of the proposed method and benchmarks.}
		\label{fulfill_ratio}
		\vspace{-.3cm}
	\end{figure}
	
	The comparison of task fulfillment ratio between benchmarks and the proposed method is shown in Fig. \ref{fulfill_ratio}. As observed, two general-purpose LLMs fail to generate valid system orchestrations in nearly all cases when applied directly. After knowledge injection and instruction fine-tuning, these models exhibit significant improvements in both output format compliance and content alignment with system requirements. However, their planning decisions remain essentially stochastic without exposure to real-world environment feedback or closed-loop training, leading to low practical success rates. By applying rejection sampling to extract high-quality positive samples and using them for supervised refinement, the model's task fulfillment ratio improves substantially. Finally, through our proposed RLDTF framework, the model achieves a near $75\%$ one-shot task completion rate, demonstrating effective learning of reliable and executable orchestration policies.

\begin{figure}
		\centering
		\includegraphics[width=0.35\textwidth]{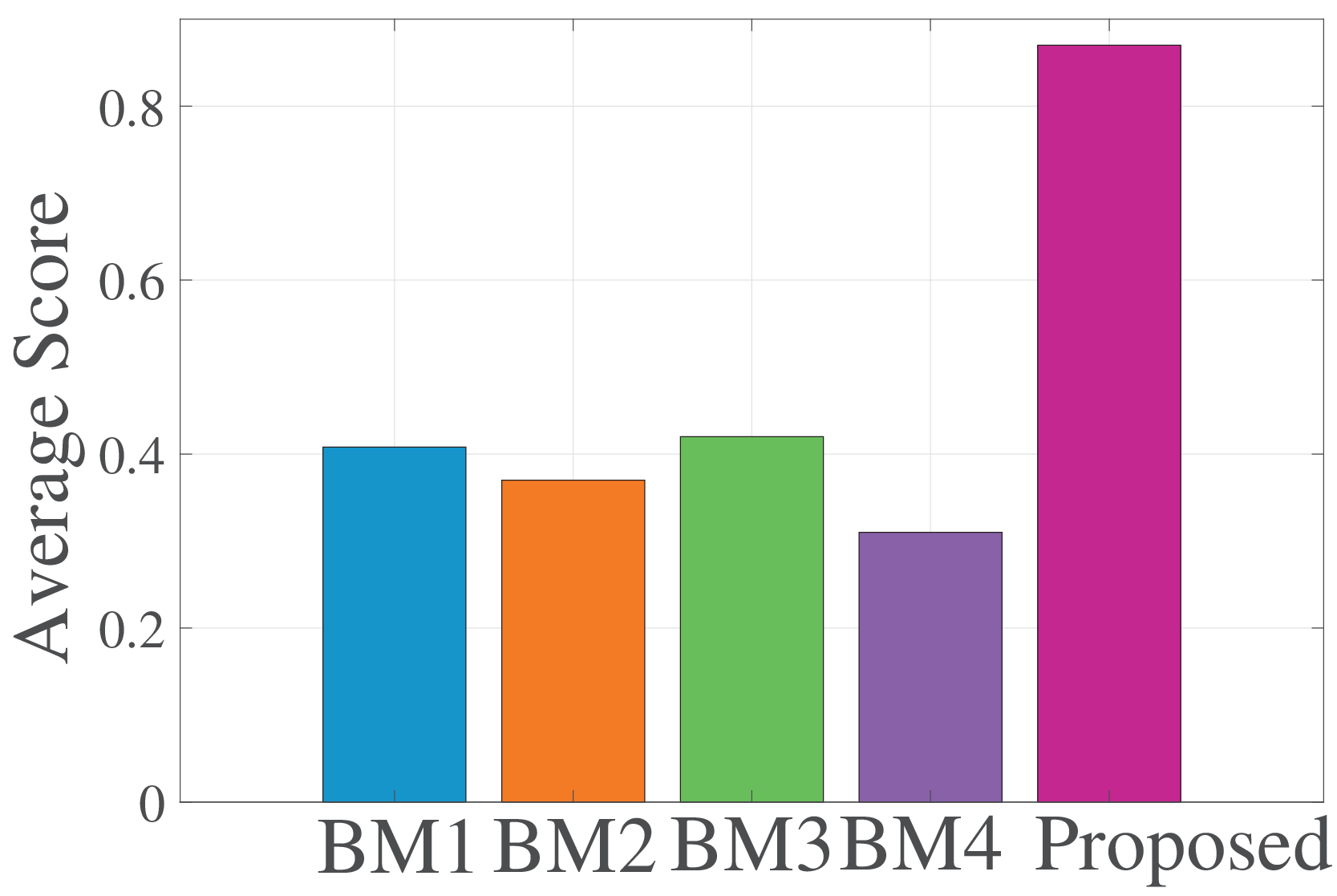}
		\vspace{-.2cm}
		\caption{Average scores for the fulfilled tasks of proposed method and benchmarks.}
		\label{average_score}
		\vspace{-.8cm}
	\end{figure}

    \addtolength{\topmargin}{0.02cm}
	Fig. \ref{average_score} further presents the average scores of responses for completed requirements across all benchmarks and the proposed method. As shown, models without reinforcement training achieve similar average scores, indicating suboptimal solution quality. While rejection sampling improves the overall fulfillment ratio by leveraging only positive examples, it introduces a bias toward resource-intensive solutions, reducing the optimality of the orchestration. In contrast, after RLDTF training, a significant increase in average score is observed. This demonstrates that RLDTF improves the likelihood of success and guides the model toward higher-quality, more efficient solutions that better align with the requirements of the upper-layer tasks.
	
	\begin{figure}
		\centering
		\includegraphics[width=0.4\textwidth]{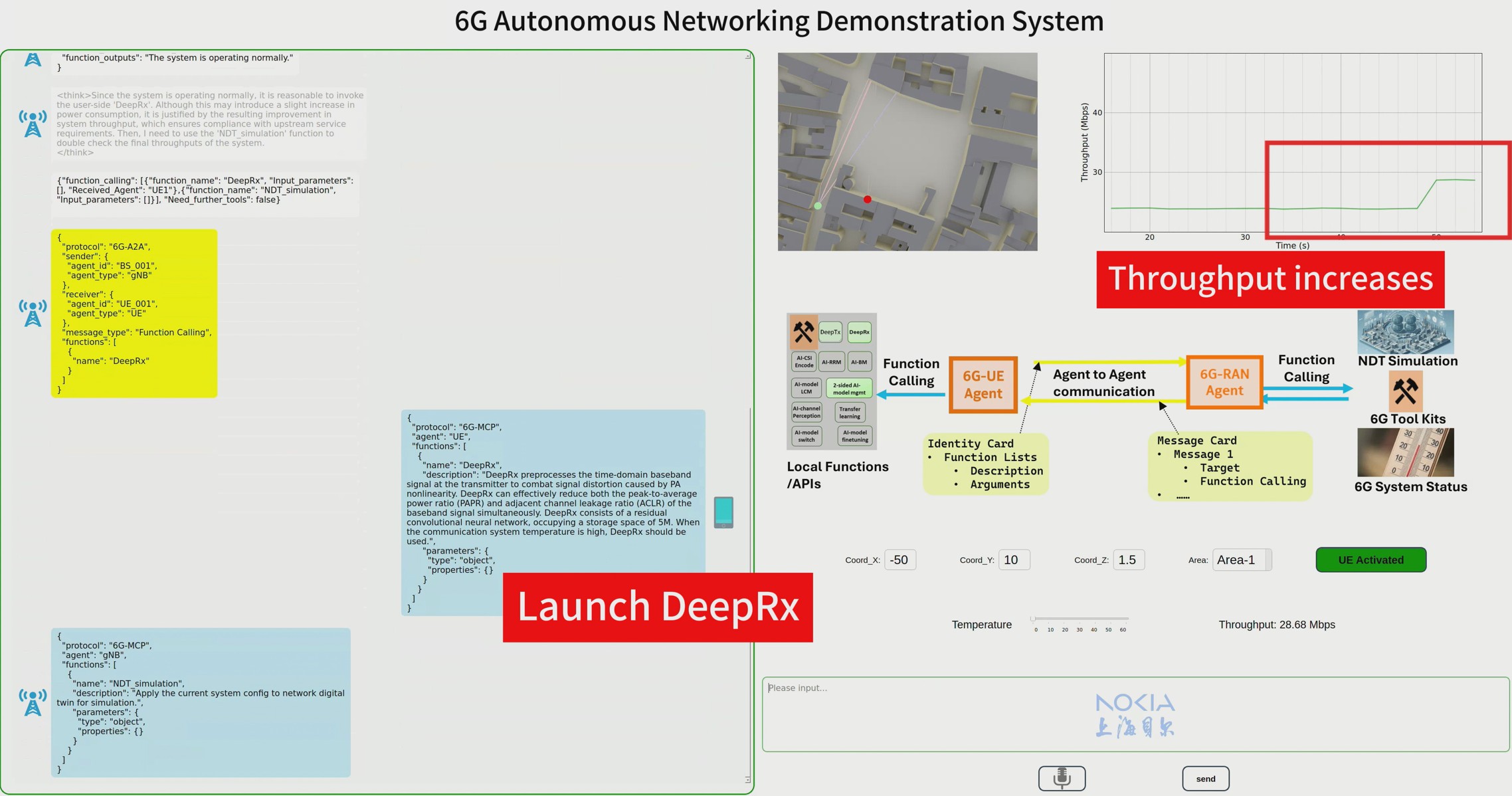}
		\caption{The live demo of 6G-LLM on real hardware}
		\label{demo}
		\vspace{-.7cm}
	\end{figure}
	
	Finally, shown in Fig. \ref{demo}, in order to intuitively illustrate the concept of applying LLMs as the 6G orchestrator and the effectiveness of our proposed methods, we have developed a real-time hardware prototype demonstration where the 6G-LLM is deployed. The demo showcases the model's ability to interpret ambiguous upper-layer task requirements and understand real-time system hardware states, then autonomously invoke and configure AI-native communication modules (e.g., DeepTx, DeepRx \cite{9345504}) as well as traditional communication algorithms, to fulfill the underlying communication needs. We sincerely hope readers will watch the illustration video, which can be obtained in '\textit{https://github.com/hongan-nokia/bell\_labs\_6G\_llm}' for further reference.
	
	\section{Conclusions} \label{conclusion}
	This paper proposed a novel RLDTF paradigm for training 6G-LLMs as intelligent orchestrators for task-oriented 6G physic-layer automation. By enabling continuous learning and dynamic optimization, RLDTF overcame the limitations of static training and scarce human-labeled data. The integration of a weighted token advantage mechanism further improved output precision. Experimental results validated the feasibility and effectiveness of the proposed methodology. We believe this work provides valuable inspiration for advancing native intelligence in future communication networks.
	
	\section*{Acknowledgment*}
	This work was supported in part by National Key Research and Development Program of China under Grant 2024YFE0200600.
	\bibliographystyle{IEEEtran}
	\bibliography{bibfile}

@ARTICLE{10700707,
  author={Long, Sifan and Tang, Fengxiao and Li, Yangfan and Tan, Tiao and Jin, Zhengjie and Zhao, Ming and Kato, Nei},
  journal={IEEE Network}, 
  title={{6G Comprehensive Intelligence: Network Operations and Optimization Based on Large Language Models}}, 
  year={2025},
  volume={39},
  number={4},
  pages={192-201},
  doi={10.1109/MNET.2024.3470774}}

@ARTICLE{10638533,
  author={Jiang, Feibo and Peng, Yubo and Dong, Li and Wang, Kezhi and Yang, Kun and Pan, Cunhua and Niyato, Dusit and Dobre, Octavia A.},
  journal={IEEE Wireless Communications}, 
  title={{Large Language Model Enhanced Multi-Agent Systems for 6G Communications}}, 
  year={2024},
  volume={31},
  number={6},
  pages={48-55},
  doi={10.1109/MWC.016.2300600}}

@ARTICLE{10648594,
  author={Xu, Minrui and Niyato, Dusit and Kang, Jiawen and Xiong, Zehui and Mao, Shiwen and Han, Zhu and Kim, Dong In and Letaief, Khaled B.},
  journal={IEEE Wireless Communications}, 
  title={{When Large Language Model Agents Meet 6G Networks: Perception, Grounding, and Alignment}}, 
  year={2024},
  volume={31},
  number={6},
  pages={63-71},
  doi={10.1109/MWC.005.2400019}}

@ARTICLE{11097898,
  author={Zou, Hang and Zhao, Qiyang and Tian, Yu and Bariah, Lina and Bader, Faouzi and Lestable, Thierry and Debbah, Merouane},
  journal={IEEE Transactions on Machine Learning in Communications and Networking}, 
  title={{TelecomGPT: A Framework to Build Telecom-Specific Large Language Models}}, 
  year={2025},
  volume={3},
  number={},
  pages={948-975},
  doi={10.1109/TMLCN.2025.3593184}}

@ARTICLE{11029511,
  author={Jiao, Tianyu and Xu, Yin and Xiao, Zhuoran and Huang, Yihang and Ye, Chenhui and Feng, Yijia and Cai, Liyu and Chang, Jiang and Liu, Fangkun and He, Dazhi and Guan, Yunfeng and Zhang, Wenjun},
  journal={IEEE Wireless Communications Letters}, 
  title={{AI2MMUM: AI-AI Oriented Multi-Modal Universal Model Leveraging Telecom Domain Large Model}}, 
  year={2025},
  volume={14},
  number={8},
  pages={2651-2655},
  doi={10.1109/LWC.2025.3578370}}

@INPROCEEDINGS{11101226,
  author={Xiao, Zhuoran and Ye, Chenhui and Hu, Yunbo and Yuan, Honggang and Huang, Yihang and Cai, Liyu and Chang, Jiang and Feng, Yijia},
  booktitle={2024 IEEE Globecom Workshops (GC Wkshps)}, 
  title={{LLM Agents as 6G Orchestrator: A Paradigm for Task-Oriented Physical-Layer Automation}}, 
  year={2024},
  volume={},
  number={},
  pages={1-6},
  doi={10.1109/GCWkshp64532.2024.11101226}}

@ARTICLE{9345504,
  author={Honkala, Mikko and Korpi, Dani and Huttunen, Janne M. J.},
  journal={IEEE Transactions on Wireless Communications}, 
  title={{DeepRx: Fully Convolutional Deep Learning Receiver}}, 
  year={2021},
  volume={20},
  number={6},
  pages={3925-3940},
  doi={10.1109/TWC.2021.3054520}}
\end{document}